\documentclass[pra,aps,reprint,nofootinbib,superscriptaddress,floatfix]{revtex4-2}

\usepackage[T1]{fontenc}
\usepackage{times}
\usepackage{graphicx}
\usepackage{epsfig}
\usepackage{amsmath,amssymb,amsfonts,amsthm}
\usepackage{dsfont}
\usepackage{bbm}
\usepackage{xcolor}
\usepackage[normalem]{ulem}
\usepackage{enumitem}
\usepackage{comment}
\usepackage{appendix}
\usepackage[colorlinks=true,linkcolor=blue,citecolor=blue,
urlcolor=blue,breaklinks]{hyperref}
\usepackage{braket}
\usepackage{physics}

\definecolor{mb}{RGB}{0,102,204}
\begin{document}
\title{Coherent Control of Channel Dilations Activate Temporal Bell Nonclassicality}

\author{H. S. Karthik} 
\email{hsk1729@gmail.com}
\affiliation{International Centre for Theory of Quantum Technologies, University of Gda\'nsk, Prof. Marii Janion 4, 80-309 Gda\'nsk, Poland.}

\author{Akshata Shenoy H.}
\affiliation{International Centre for Theory of Quantum Technologies, University of Gda\'nsk, Prof. Marii Janion 4, 80-309 Gda\'nsk, Poland.}

\author{A. R. Usha Devi}
\affiliation{Department of Physics, Bangalore University, Bengaluru-560 056, India}

\author{Marcin Paw\l{}owski}
\affiliation{International Centre for Theory of Quantum Technologies, University of Gda\'nsk, Prof. Marii Janion 4, 80-309 Gda\'nsk, Poland.}

\begin{abstract}
The temporal Clauser–Horne–Shimony–Holt (CHSH) inequality witnesses the nonclassicality of temporal correlations, but its violation is generally degraded by environmental noise. Here, we show that violation of the temporal CHSH inequality can be revived through coherent control of noisy quantum evolutions. We compare two physically distinct implementations: coherent control of noisy evolutions induced by interaction of the system with independent environments, and coherent control of two physically distinct, unitarily equivalent Stinespring dilations of the same noisy channel. Although these constructions generate identical deterministic system dynamics, they induce distinctly different post-selected evolutions. With a focus on the amplitude-damping channel (ADC), we show that coherent control of equivalent dilations extend the range of temporal CHSH inequality violation well beyond both the incoherently controlled, or deterministic, scenario and what is achievable with independent environments. Under setting-independent post-selection of the coherent control implementation, the resulting violation further certifies that the channel is not strongly CHSH nonlocality-breaking. Our results identify the choice of Stinespring dilation as an operationally relevant resource in coherently controlled tests of temporal quantum correlations.
\end{abstract}

\maketitle

\section{Introduction}
Correlations between physical observables within quantum theory (QT) exhibit intriguing features fundamentally different from classical physics. Quantum correlations call out for explanation \cite{Bell1} irrespective of whether they manifest in spatially separated bipartite systems or from  sequential measurements performed on a single system. A particularly interesting point of enquiry is exploring the quantum-classical boundary in single systems through temporal correlations arising from sequential measurements. 

Two complementary \textit{operational} implementations are commonly employed to witness the nonclassicality of temporal correlations: the temporal Clauser–Horne–Shimony–Holt (CHSH) inequality \cite{Brukner2004,Fritz2010}, which is an analogue of the spatial CHSH inquality \cite{chsh}, and the Leggett–Garg inequality (LGI), which is a linear correlation inequality constructed out of measurements of a single observable at multiple instants of time. Both the inequalities can be derived based on the postulates of Macrorealism (MR), viz.,(i) Macrorealism per se: Physical properties of a macroscopic system exist independent of the act of observation and (ii) Non-invasive measurability: Measurements performed at an instant do not influence the subsequent system evolution \cite{LG85,emary+14,halliwell2016leggett, halliwell2017comparing, zhang2023comparing}. Violations of both the temporal CHSH and the LGI reveal incompatibility of QT with these classical assumptions. Together, these approaches provide complementary routes for investigating the quantum–classical boundary through sequential measurements on single systems \cite{Fritz2010,KB13,zhang2023comparing}.

Violations of LGI have been experimentally demonstrated across a wide range of physical platforms, including superconducting circuits \cite{palacios2010experimental}, nuclear-spin systems \cite{PhysRevLett.107.130402,PhysRevA.87.052102}, nitrogen-vacancy centers in diamond \cite{PhysRevLett.107.090401}, photonic systems \cite{PhysRevLett.106.040402,PhysRevLett.115.113002}, trapped ions \cite{PhysRevA.107.012424} and single-spin solid-state platforms \cite{PhysRevA.105.042613}.    
Experimental violations of temporal Bell inequalities have likewise been observed in photonic platforms using sequential measurements \cite{Fedrizzi2011} and in single-spin solid-state systems based on nitrogen-vacancy centers in diamond \cite{Waldherr2011}. Together, these investigations establish temporal-correlation inequalities as experimentally accessible probes of quantum behavior and provide direct evidence for the nonclassical nature of correlations across time.

Beyond their foundational significance, temporal quantum correlations, have increasing relevance in witnessing the dimension of quantum systems \cite{spee2020genuine}, certifying and benchmarking quantum memories \cite{rosset2018resource} and, enhancing performance of quantum clocks \cite{budroni2021ticking}. They also provide powerful tools for detecting non-Markovian behavior through coherence-based approaches \cite{wu2020detecting} and for characterizing multi-time quantum processes \cite{berk2021resource}. Furthermore, temporal Bell-type inequalities arising from sequential measurements have highlighted the role of non-classical correlations in time as a resource for quantum information processing \cite{Zukowski2014Temporal}. Other applications of temporal correlations include self-testing of quantum measurements \cite{maity2021self} and certifying quantum channels within minimal-assumptions in device-independent scenarios \cite{pusey2015verifying}. 

A major challenge in observing the quantum nature of temporal correlations is their fragility to environmental noise. Decoherence and dissipation effects suppress such correlations, leading to critical noise thresholds beyond which, quantumness disappears \cite{emary+14}. For example, the presence of environmental noise between temporally separated measurements, rapidly weakens the violation of LGI (also the temporal CHSH inequality), thereby limiting the witnessing of temporal quantum effects in realistic experimental scenarios \cite{LG85, PhysRevLett.107.130402, knee2016strict, huffman2017violation}.

In scenarios involving spatial correlations, quantum resources such as entanglement \cite{ruskai2003qubit, horodecki2003entanglement}, steering \cite{RevModPhys.92.015001,du2025dynamics, srivastav2022quick}, and Bell nonlocality \cite{RevModPhys.86.419, pal2015non} degrade in the presence of noise. Nevertheless, through stochastic local filtering \cite{kim2012protectingentanglement,PhysRevLett.111.160402,PhysRevA.99.030101,PhysRevA.109.022411}, the nonclassical  nature of the respective quantum correlations can be revived beyond certain noise thresholds. Analogously, temporal nonclassicality too can be activated using stochastic operations. In particular, it was demonstrated in \cite{KWS+21} that suitably chosen filtering operations can revive violations of the temporal CHSH inequality even when deterministic evolution fails to reveal nonclassical behavior, a phenomenon termed as \emph{hidden nonmacrorealism}. Similar ideas have been employed for quantum communication where probabilistic strategies activate the quantum advantage of quantum random access codes against noise \cite{KGQ+25}. These results demonstrate that post-selection on successful stochastic filtering events, performed before the relevant measurements, can reveal quantum features that remain hidden in conventional deterministic protocols.

A complementary strategy for combating the noise-induced degradation of quantum signatures is provided by coherent control of quantum processes. In coherently controlled dynamics, a quantum control system places alternative evolutions in superposition, enabling operational advantages that are unattainable through classical mixtures of quantum channels \cite{Oi03,LongGui-Lu_2006, CK19,Abbott2020Communication, pang2023experimental}. Such approaches, based on coherent control and process engineering, further demonstrate that quantum coherence can be systematically harnessed to realize optimal quantum communication tasks beyond conventional channel paradigms \cite{garciadiaz2018using, RRE+21}. Pertinently, these ideas have found applications in  noise mitigation \cite{RRE+21}, duality quantum computing \cite{LongGui-Lu_2006, LongGui-Lu_2008} including recent demonstrations of steering-assisted metrology using superpositions of noisy phase-shifts \cite{LLM+23}. More broadly, channel-based approaches have also been employed for the certification of quantum memories \cite{yuan2021universal}. Furthermore, coherent superpositions of unitary evolutions have recently been shown to yield extreme violations of LGI while exhibiting enhanced robustness against noise (acting either on the system or the control qubit), revealing a strong connection between coherent control and temporal quantum correlations \cite{CKM+25}.

This connection becomes particularly important when the noise acting on the system is considered at the level of its physical implementation. It is well known that different Stinespring dilations correspond to the same reduced noise channel when the evolution is used deterministically. However, under coherent control, such implementation-level distinctions need not remain operationally hidden. Specifically, coherently controlling different realizations of the same reduced noise channel can lead to distinctly different conditional dynamics, even when the corresponding deterministic evolution is identical. The implications of this implementation dependence for temporal quantum correlations remain unexplored. 

In this work, we address this question by investigating how coherent control and post-selection activate the violation of the temporal CHSH inequality in an amplitude-damped system qubit. Specifically, we compare two physically distinct realizations of coherently controlled noisy evolution. The first corresponds to coherent control of noisy evolutions induced by interactions between the system and independent environments, while the second concerns coherent control of two physically distinct, unitarily equivalent Stinespring dilations of the ADC. Although both constructions realize the same ADC in the absence of coherent control, we show that they generate distinctly different post-selected evolutions. Consequently, the extent to which temporal CHSH inequality violation can be activated differs significantly between these two scenarios. In particular, coherent control at the dilation level enables activation over a broad range of noise strengths and extends the noise threshold for violation far beyond that attainable using independently controlled noise channels. Our results therefore establish that the choice of Stinespring dilation can acquire direct operational significance in coherently controlled tests of temporal correlations and reveal a novel mechanism for preserving and enhancing the quantum nature of temporal correlations in noisy scenarios.

The rest of this paper is organized as follows. In Sec.~\ref{sec:pre}, we review the preliminary notions related to the temporal-CHSH inequality and the ADC. In Sec.~\ref{sec:ccc}, we discuss the two implementations of coherently controlled noisy evolutions. At first, we review the coherent control of channels induced by the interaction between the system and independent environments and then introduce the novel coherent control of unitarily equivalent Stinespring dilations. The corresponding results specific to ADC are also presented here.  We present the main results of the paper concerning the activation thresholds obtained under the two implementations in Sec.~\ref{Sec:at}. Finally in, Sec.~\ref{sec:dis} we discuss and summarize the results. A concise discussion on the connection of Choi matrix, non-locality breaking channels, correspondence with temporal CHSH inequality activation is discussed in Appendix~\ref{app:cm}.

\section{Preliminary notions}
\label{sec:pre}
\subsection{Temporal-CHSH inequality}
\label{subsec:tchsh}
Under the assumptions of MR, the two-time joint probability distribution can be expressed as
\begin{equation}
P(a,b|A_i,B_j)_{\mathrm{MR}} =\sum_{\lambda} p(\lambda)~
p(a|A_i,\lambda)~ p(b|B_j,\lambda),
\label{eq:MR}
\end{equation}
where $P(a,b|A_i,B_j)$ denotes the probability of obtaining outcomes $a$ and $b$ ($a,b\in{\pm1}$) from sequential measurement of observables $A_i$ and $B_j$, performed at earlier and later times, respectively, with $i,j\in{1,2}$. The hidden variable $\lambda$ predetermines the outcomes through the conditional probability distributions $p(a|A_i,\lambda)$ and $p(b|B_j,\lambda)$, while $p(\lambda)$ is the probability distribution over the hidden variable.

For a given pair of measurements $(A_i,B_j)$, the temporal correlation is defined as
\begin{equation}
C_{ij}= \sum_{a,b=\pm1} ab\,P(a,b|A_i,B_j).
\label{eq:corr}
\end{equation}

Using these correlation functions, the temporal CHSH parameter is given by
\begin{equation}
S_T = C_{11}+C_{12}+C_{21}-C_{22},
\label{eq:ST}
\end{equation}
which satisfies the macrorealistic bound $S_T\leq 2$ \cite{Fritz2010}.

Now consider a single qubit subjected to two sequential measurements separated by an intermediate dynamical evolution described by a quantum channel $\mathcal{E}$, such that $\rho \mapsto \mathcal{E}_{\gamma}(\rho)$. In QT, the corresponding two-time joint probability is written as
\begin{equation}
P(a,b|A_i,B_j)_{\mathrm{QT}} =\mathrm{Tr}\left[ M_{b|B_j}\mathcal{E}\left(
M_{a|A_i}\rho_0 M_{a|A_i}^{\dagger}\right)\right],
\label{eq:QMprob}
\end{equation}
where $\rho_0 \equiv \rho_S(0)$ is the initial state of the qubit (system qubit). $M_{a|A_i}$ and $M_{b|B_j}$ denote the projectors corresponding to measurement observables $A_i$ and $B_j$ respectively. For a maximally mixed initial state, $\rho_0 = \frac{\mathbb{I}}{2}$, Eq.~\eqref{eq:QMprob} reduces to
\begin{equation}
P(a,b|A_i,B_j) =\frac{1}{2} \mathrm{Tr} \left[ M_{b|B_j} \mathcal{E}(M_{a|A_i})
\right].
\label{eq:channelprob}
\end{equation}

Unlike the macrorealistic description of Eq.~\eqref{eq:MR}, QT predicts temporal correlations that can violate the temporal CHSH inequality, yielding $S_T > 2$. Such a violation demonstrates that the quantum two-time joint probability distribution cannot, in general, be decomposed into the Eq.~\eqref{eq:MR} form. For the identity channel, i.e, when $\mathcal{E} = \mathbb{I}$, choosing the measurement operator set 
\begin{align}
A_1 &= X, &
B_1 &= \frac{X+Y}{\sqrt{2}} \nonumber\\ 
A_2 &= Y, &
B_2 &= \frac{X-Y}{\sqrt{2}},
\label{eq:xymeas}
\end{align}
the maximum attainable quantum violation of $2\sqrt{2}$ can be reached, which is known as the Temporal Tsirelson bound (TTB). 

\subsection{Amplitude Damping Channel}
\label{subsec:adc}
Amplitude damping provides a classic model of dissipation  from the excited state to the ground state for a qubit undergoing irreversible relaxation dynamics and is known to rapidly suppress temporal quantum correlations. 

An ADC with a damping parameter $\gamma\in[0,1]$ is described by the Kraus operators
\begin{eqnarray}
\label{eq:k0}
K_0 &=& |0\rangle\!\langle0|
      +\sqrt{1-\gamma}\,|1\rangle\!\langle1|,\\
\label{eq:k1}
K_1 &=& \sqrt{\gamma}\,|0\rangle\!\langle1|.
\end{eqnarray}
Acting between sequential measurements, an ADC progressively suppresses temporal quantum correlations leading to weakened violation of the temporal CHSH inequality up to a threshold damping strength beyond which, the violations of the inequality vanishes.

\medskip
\emph{System--environment dilation for the ADC:} Any \textit{completely positive and trace preserving} (CPTP) map admits a Stinespring dilation~\cite{NielsenChuang}, in which the channel is realized through a unitary interaction between the system and an environment initially prepared in a fixed state. Accordingly, an ADC can be represented as,
\begin{equation}
\mathcal{E}(\rho_S) = \operatorname{Tr}_{E} \!\left[ U_{SE}
\left( \rho_S\otimes |0\rangle\!\langle0|_E \right) U_{SE}^{\dagger} \right],
\label{eq:stinespring}
\end{equation}
where $U_{SE}$ is a unitary acting on the joint system--environment Hilbert space, $\rho_S$ is the state of the system and  $|0\rangle\!\langle0|_E$ is the initial state of the environment.

Since we are going to discuss about coherently controlled implementation of the noisy evolutions, instead of focusing solely on its reduced CPTP description, it is natural to work within a system--environment dilation of the ADC. Considering, as before a system qubit $S$ interacting with an environmental qubit $E$, initially prepared in the state $|0\rangle_E$, the corresponding Stinespring dilation is generated by the Hamiltonian
\begin{equation}
    H_{SE} = g\,G = \frac{g}{2} \left( \sigma_x^{(S)}\!\otimes\sigma_y^{(E)} -
\sigma_y^{(S)}\!\otimes\sigma_x^{(E)}
\right),
\end{equation}
where $g$ denotes the interaction strength and
$G~:=~i\Big(|01\rangle\!\langle10|~-~|10\rangle\!\langle01| \Big)$
is the corresponding Hermitian generator. Define $\theta:=gt$,
and express the amplitude-damping parameter (see Eqs.~(\ref{eq:k0}) and (\ref{eq:k1}))   by 
\begin{equation}
\gamma(\theta) = \sin^2\theta.
\end{equation}
Given this, a particular realization of the resulting joint unitary evolution can be expressed as 
\begin{equation}
U_{SE}(\theta) = e^{-i\theta G} =
\begin{pmatrix}
1 & 0 & 0 & 0\\
0 & \cos\theta & \sin\theta & 0\\
0 & -\sin\theta & \cos\theta & 0\\
0 & 0 & 0 & 1
\end{pmatrix},
\end{equation}
written in the computational basis
$\{|00\rangle,|01\rangle,|10\rangle,|11\rangle\}$.

\medskip

Considering the evolution of a qubit between successive measurements according to Eqs.~(\ref{eq:k0}) and (\ref{eq:k1}), 
\begin{equation}
\rho \mapsto
\mathcal{E}_{\gamma}(\rho) = K_0 \rho K_0^\dagger +
K_1 \rho K_1^\dagger.
\label{eq:adc}
\end{equation} 
Consequently, substituting the measurement set of Eq.~\eqref{eq:xymeas} in Eq.~\eqref{eq:channelprob} for the ADC  yields the two-time correlations (see Eq. \eqref{eq:corr}). From this, the temporal CHSH inequality can be evaluated as   $\mathcal{S}_T~=~2\sqrt{2(1 - \gamma)}$. This expression shows that $S_T$ decreases monotonically with increasing damping strength $\gamma$, implying a weakening of the temporal quantum correlations due to environmental noise. Furthermore, beyond a critical damping parameter $\gamma_c = 0.5$, the violation of the temporal CHSH inequality completely vanishes.

Under the Choi isomorphism  \cite{Jamiolkowski1972,Choi1975,NielsenChuang}, temporal CHSH and spatial CHSH correlations share an identical mathematical description, since both are determined by a common correlation tensor. The Choi state  (see Appendix \ref{app:cm}) encodes the quantum channel governing the temporal evolution between sequential measurements on a single system. In the spatial CHSH scenario, the  Choi state is interpreted as a bipartite state on which local measurements are performed. The distinction between the two scenarios therefore lies in their physical realization rather than the mathematical structure.

The Stinespring dilation in Eq. (\ref{eq:stinespring}) provides a natural framework for implementing coherent control at the level of channel realizations. Whereas a channel encoded by the Choi state can always be realized as a unitary interaction between a system and an environment, we now investigate two physically distinct coherently controlled implementations of noise evolutions. 

\section{Two Coherent Control Implementations}
\label{sec:ccc}
We now consider two physically distinct implementations of coherently controlled noisy evolutions. The first implementation coherently controls channel realizations arising due to coupling the system with two independent environments, while the second, coherently controls two unitarily equivalent Stinespring dilations of a single channel. Although both implementations induce an identical reduced channel on the system, they give rise to distinct conditional dynamics.

\subsection{Coherent Control of Quantum Channels}
\label{subsec:soc}
\begin{figure*}[ht!]
\centering
\includegraphics[width=0.9\textwidth]{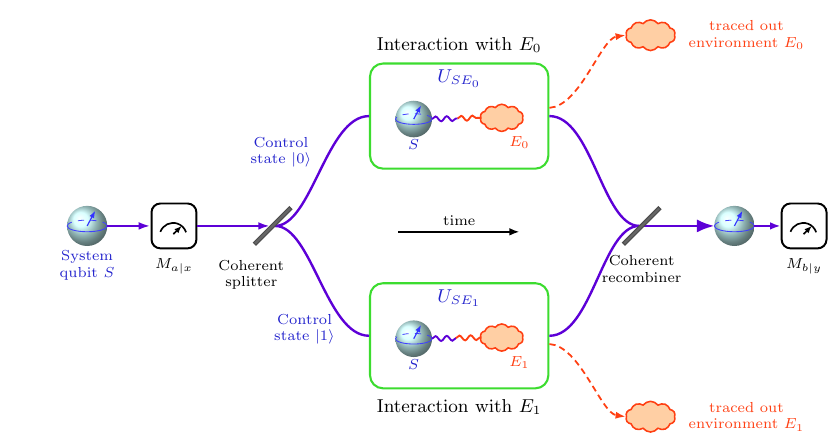}
\caption{Schematic of a temporal CHSH inequality experiment using sequential measurements and coherently controlled interaction of the system with independent environments: First, an initial quantum state is measured. Next, the state undergoes noisy evolution resulting from the unitary interaction with independent environments $E_1$ and $E_2$. A control qubit C (acting as the which path qubit) coherently controls the choice of this interaction. After the coherent evolution, the environments are traced out and the branches are coherently recombined (control post-selection and trace out). Finally, the effectively evolved system undergoes a second measurement. Time runs from left to right.}
\label{fig:supind}
\end{figure*}  
In the first implementation, a control qubit coherently selects between two realizations of the same noisy evolution, each arising from the coupling of the system with a distinct environment. The environments, denoted by $E_0$ and $E_1$, are independent and initially uncorrelated. This setup corresponds to the standard notion of coherent control of quantum channels or \emph{superposition of channels} \cite{Oi03, Abbott2020Communication, pang2023experimental}, where each control state is associated with a distinct channel implementation (see Fig.\ref{fig:supind}). For the sake of comparison and completeness, let us briefly review this framework below. \\

\emph{Global controlled unitary:}
We introduce a control qubit $C$, a system qubit $S$, and two  environment systems
$E_0$ and $E_1$, each initially prepared in the same state.
The global unitary acting on the combined Hilbert space $C\otimes S\otimes E_0\otimes E_1$ is defined as 
\begin{equation}
W^\mathrm{ind} = |0\rangle\!\langle 0|_{C}\otimes U_{0, S E_0} +
|1\rangle\!\langle 1|_{C}\otimes U_{1, S E_1},
\end{equation}
where $U_{0, S E_0}$ and $U_{1, S E_1}$ represents the two  system--environment unitaries associated with the two controlled branches and `ind' denotes that, in each branch, the system couples to independent and distinct environment  \cite{Oi03,Abbott2020Communication}.

Let the initial joint state of the control, system, and environments be
\begin{equation}
\rho^\mathrm{ind}_{\mathrm{tot}}(0)
=
\rho_C(0)\otimes \rho_S(0)
\otimes |0\rangle\!\langle0|_{E_0}
\otimes |0\rangle\!\langle0|_{E_1},
\end{equation}
with $\rho_C(0) = |q\rangle\!\langle q|_C$, where `q' can be either 0 or 1.\\

\emph{Joint state and conditional system states:} After the controlled evolution, the total state evolves to
\begin{equation}
\rho^\mathrm{ind}_{\mathrm{tot}} =
W^\mathrm{ind}~\rho^\mathrm{ind}_{\mathrm{tot}}(0)~ {W^\mathrm{ind}}^\dagger,
\end{equation}
from which the reduced control-system state can be obtained after tracing out the environments $E_0$ and $E_1$. That is, 
\begin{eqnarray}
\rho^\mathrm{ind}_{CS} &=& \operatorname{Tr}_{E_0,E_1} \!\left[
\rho^\mathrm{ind}_{\mathrm{tot}} \right] \nonumber \\
&=& |q\rangle\langle q|_C  \otimes \mathcal{E}_{q}(\rho_S), 
\end{eqnarray}
where $\mathcal{E}_{q}(\rho_S)$ is the channel realized after interaction with environment selected by the value of `q'. In other words, when the control C is initialized in the state $\ket{q}$, the system S undergoes the transformation through the channel realization $U_{q,SE_q}$.
 
To this end, if C is prepared as a maximally incoherent state, i.e., $\rho_C(0)=\frac{(\ket{0}\bra{0}_C + \ket{1}\bra{1}_{C})}{2}$, the system S evolves either through the unitary interacting with $E_0$ or $E_1$ with probability $\frac{1}{2}$. 

On the other hand, if the control qubit is initially prepared in the (maximally coherent) state $\rho_C(0)=|\pm\rangle\!\langle\pm|_{C},$  $|\pm\rangle{_C}~=~\frac{(|0\rangle_{C}\pm|1\rangle_{C})}{\sqrt{2}}$, then after the controlled evolution and tracing out the environments, the joint control-system state becomes,
\begin{equation}
\begin{aligned}
\rho ^\mathrm{ind}_{CS} =& \frac{1}{2} \Big(|0\rangle\!\langle0|_{C} \otimes \mathcal{E}_0(\rho_S) + |1\rangle\!\langle1|_{C} \otimes \mathcal{E}_1(\rho_S)\Big) \\
&\pm \frac12 \Big(|0\rangle\!\langle1|_C \otimes T_0 \rho_S T^\dagger_1 + |1\rangle\!\langle0|_C \otimes T_1 \rho_S T^\dagger_0\Big),
\end{aligned}
\label{eq:rhoCS_ind}
\end{equation}
where 
\begin{equation}
T_i = \rm{Tr}_{E_i} \!\left[U_{i,SE_i} (\mathbb I_S\otimes\rho_{E_i}) \right] = {}_{E_i}\langle0|U_{i,SE_i} |0\rangle_{E_i}
\label{eq:Toperator}
\end{equation}
is the interference operator governing the coherence between the two controlled independent environment branches. $\rho_{\mathrm {E_i}}$ is the initial state of the environment $i$. 

Now, projective measurement of the control qubit in the Hadamard basis
$\{|\pm\rangle_{_C}\}$, with projectors $\Pi_{C,\pm}=|\pm\rangle\!\langle\pm|_{C}$,
yields the unnormalized conditional system state as
\begin{equation}
\begin{aligned}
\tilde{\rho}^\mathrm{ind}_{S,\pm}
&= \operatorname{Tr}_C \!\Big[(\Pi_{C,\pm}\otimes\mathbb I_S)
\rho_{CS}^\mathrm{ind} \Big] \\
&= \frac{1}{4} \Big(\mathcal{E}_0(\rho_S) + \mathcal{E}_1(\rho_S)\Big)  \pm \frac{1}{4} \Big(T_0 \rho_S T^\dagger_1 + T_1 \rho_S T^\dagger_0\Big).
\end{aligned}
\label{eq:conditional_ind1}
\end{equation}

Equivalently, we can define the \textit{effective} system–environment operators associated with the two control outcomes $\pm$ as,

\begin{equation}
A_{\pm}^{\mathrm{ind}} = \frac{1}{2}  \left( U_{0, SE_0} \otimes \mathbb{I}_{E_1}~\pm~ U_{1, SE_1} \otimes \mathbb{I}_{E_0} \right),
\end{equation}
where the $U_{0, SE_0} \otimes \mathbb{I}_{E_1}, U_{1, SE_1} \otimes \mathbb{I}_{E_0}$ denote their embeddings on the joint Hilber space $(S \otimes E_0 \otimes E_1)$ with the identity acting on the inactive environment in that branch. 
The corresponding unnormalized \textit{conditional} channel can then be expressed as,
\begin{equation}
\widetilde{\mathcal{E}}_{\pm}^{\mathrm{ind}}(\rho_S)
= \operatorname{Tr}_{E_0E_1} \left[A_{\pm}^{\mathrm{ind}}
\left(\rho_S\otimes |00\rangle\langle00|_{E_0E_1} \right) A_{\pm}^{\mathrm{ind}\dagger} \right].
\end{equation}

Expanding this expression, we obtain
\begin{align}
\widetilde{\mathcal{E}}_{\pm}^{\mathrm{ind}}(\rho_S)
&= \frac{1}{4} \left[ \mathcal{E}_0(\rho_S)+\mathcal{E}_1(\rho_S) \right] \nonumber\\
&\quad \pm \frac{1}{4} \left[ T_0\rho_S T_1^\dagger + T_1\rho_S T_0^\dagger \right].
\label{eq:ind-map}
\end{align}

For simplicity, let us now consider the case in which the two implementations (i.e, from $U_{0, SE_0}$ or $U_{1, SE_1}$) result in the same reduced channel on the system, that is,
\begin{equation}
\mathcal{E}_0(\rho_S) = \mathcal{E}_1 (\rho_S) \equiv \mathcal{E}(\rho_S).
\label{eq:identical_channels}
\end{equation}
The unnormalized conditional system state, then simplifies to, 

\begin{equation}
\tilde{\rho}^\mathrm{ind}_{S,\pm}
= \frac{1}{2} \mathcal{E}(\rho_S) \pm \frac{1}{4} \Big(T_0 \rho_S T^\dagger_1 + T_1 \rho_S T^\dagger_0\Big).
\label{eq:conditional_ind2}
\end{equation}

Consequently, in the case of two identical implementations, for which $T_0=T_1=T$, Eq.~(\ref{eq:conditional_ind2}) further reduces to, 

\begin{equation}
\tilde{\rho}^\mathrm{ind}_{S,\pm} = \frac{1}{2} \mathcal{E}(\rho_S) \pm \frac{1}{2} T \rho_S T^\dagger.
\label{eq:conditional_ind3}
\end{equation}
Equivalently, the corresponding unnormalized conditional channel too gets simplified to the form, 
\begin{equation}
\widetilde{\mathcal{E}}_{\pm}^{\mathrm{ind}}(\rho_S)
= \frac{1}{2}\mathcal{E}(\rho_S) \pm \frac{1}{2}T\rho_S T^\dagger.
\label{eq:identical-indenv}
\end{equation}
The post-selection probabilities can now be computed using
\begin{equation}
P^\mathrm{ind}_\pm = \operatorname{Tr} \big[ \tilde{\rho}^\mathrm{ind}_{S,\pm} \big],
 \end{equation}
and the normalized conditional states are
\begin{equation}
\rho^\mathrm{ind}_{S,\pm} = \frac{\tilde{\rho}^\mathrm{ind}_{S,\pm}}{P^\mathrm{ind}_\pm}.
\end{equation}
Thus, the states $\rho^\mathrm{ind}_{S,\pm}$ define the effective conditional dynamics associated with coherent control of the noisy evolutions implemented using independent environments. We now specialize these expressions to the amplitude-damping dilation introduced in Sec.~(\ref{subsec:adc}) and derive the corresponding conditional states explicitly.

\medskip

\emph{Specialization to ADC:}  For the ADC with $\rho_E=|0\rangle\langle0|$, $T(\theta) := \mathrm{Tr}_E[U_{SE}(\theta)(\mathbb{I} \otimes \rho_E)] = \bra{0}_E (U_{SE}(\theta))\ket{0}_E$.
Hence, $T(\theta) = K_0(\theta)$.
The unnormalized states can now be written as (see Eq. \eqref{eq:conditional_ind3}), 
\begin{equation}
\tilde{\rho}^\mathrm{ind}_{S,\pm}(\theta) =
\frac12\,\mathcal{E}_\theta(\rho_S)\pm
\frac{1}{2}K_0(\theta)\rho_S K_0^\dagger(\theta).
\end{equation}

To this end, we now have two clear ways to check activation in this superposition of amplitude-damping channel scenario:
\begin{enumerate}
    \item Prepare the control in a maximally incoherent state, i.e., $\rho_C(0)= \frac{(\ket{0}\bra{0}_{C} + \ket{1}\bra{1}_{C})}{2}$. 
\end{enumerate}
    \noindent In this case, the joint state of the control-system after the controlled evolution is given as $\frac{\ket{0}\bra{0}_{C}~+~\ket{1}\bra{1}_{C}}{2}~\otimes~ \mathcal{E_\theta} (M_{a|A_i}\rho_0 M_{a|A_i}^{\dagger})$
    where the two distinct branches implement the same ADC. Now, projecting  the control in the Hadamard basis $\{|\pm\rangle\}$, with projectors $\Pi_\pm=|\pm\rangle\!\langle\pm|$ and tracing it out leads to an updated unnormalized state of the system which is proportional to $\mathcal{E_\theta} (M_{a|A_i}\rho_0 M_{a|A_i}^{\dagger})$. After normalization, this is identical to the state obtined using a single amplitude damping channel. Thus, when $\rho_0 = \frac{\mathbb{I}}{2}$, we recover $P(a,b|A_i,B_j) =\frac{1}{2} \mathrm{Tr} \left[ M_{b|B_j} \mathcal{E}(M_{a|A_i})\right]$, which is exactly the expression for a single noise channel. Hence, we can infer that the violation of temporal CHSH inequality is not activated beyond the usual threshold of the ADC $\gamma_c=0.5$. 
\begin{enumerate}[resume]
    \item Prepare the control in the coherent state $\rho_C(0)~=~\ket{+}\bra{+}_C$.
\end{enumerate}    
    \noindent The post-measurement state $M_{a|A_i}\rho_0 M_{a|A_i}^{\dagger}$ and the control state are  evolved jointly, after which the control state is projected onto, $\Pi_\pm=|\pm\rangle\!\langle\pm|_{C}$, and traced out. For $\rho_0 = \frac{\mathbb{I}}{2}$, the corresponding (unnormalized) two-time probability becomes, 
    \begin{equation}
    \widetilde{P}_{\pm}(a,b|A_i,B_j) =\frac{1}{2} \mathrm{Tr} \left[ M_{b|B_j} \widetilde{\mathcal{E}}_{\pm}^{\mathrm{ind}}(M_{a|A_i})\right].
    \label{eq:indprob}
    \end{equation} 
    The success probability of the post-selected control measurement is given by $P^\mathrm{ind}_\pm = \operatorname{Tr} \big[ \tilde{\rho}^\mathrm{ind}_{S,\pm} \big]$. The normalized probability distribution can then be calculated as $\frac{\widetilde{P}_{\pm}(a,b|A_i,B_j)}{P^\mathrm{ind}_\pm}$. To this end, the interference terms in $\widetilde{\mathcal{E}}_{\pm}^{\mathrm{ind}}(M_{a|A_i})$ aid in activation of the temporal CHSH inequality violation beyond the threshold obtained for incoherently controlled ADCs (see Appendix \ref{app:cm} for the formulation in terms of the associated Choi state).
    
\medskip

\begin{figure}[!htbp]
\centering
\includegraphics[width=0.5\textwidth]{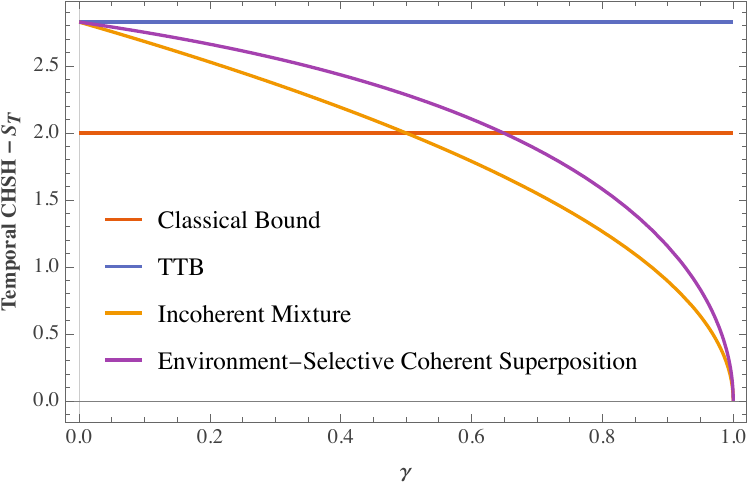}
\caption{The temporal CHSH inequality as a function of the damping parameter $\gamma$: The red and the blue lines represent the classical bound and the temporal Tsirelson bound (TTB) respectively. The orange curve corresponds to the case where the action of the environments are incoherently controlled. The purple curve represents the coherent control of the channel implementations using independent environments. It is observed that the violation of the temporal CHSH inequality is activated beyond the incoherently controlled (or the deterministic) ADC threshold and persists up to $\gamma \approx 0.65$.}
\label{fig:ind-tchsh}
\end{figure}

Thus, although post-selection induces a non trace-preserving map on the system, interference between the two channel implementation is limited by the presence of distinct environmental records. Numerically, we find that temporal CHSH inequality violation can be activated beyond the incoherently controlled ADC threshold, but only up to damping strength $\gamma \approx 0.65$ as shown in Fig. \ref{fig:ind-tchsh}.

\subsection{Coherent Control of Stinespring Dilations}
\label{subsec:sod}
\begin{figure*}[ht!]
\centering
\includegraphics[width=1.0\textwidth]{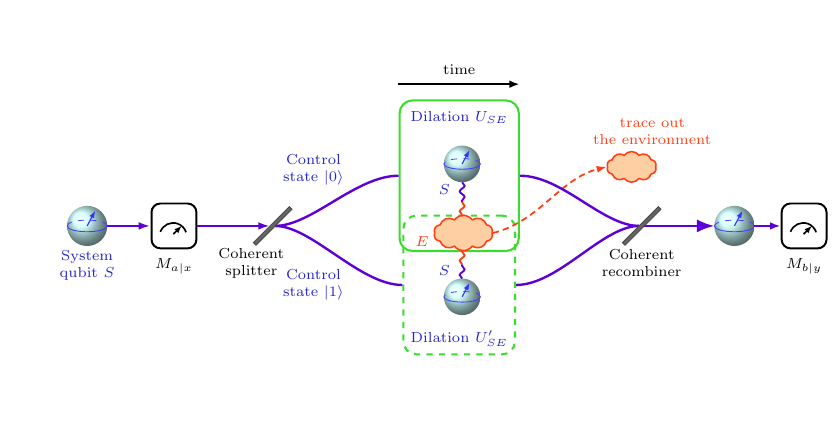}
\caption{Schematic of a temporal CHSH inequality experiment using sequential measurements and coherently controlled Stinespring dilations of the noise channel acting on the system : First, an initial quantum state is measured. Following which, the state undergoes noisy evolution resulting from the unitary interaction with an environment $E$. A control qubit C (acting as the which path qubit) coherently controls the choice of Stinespring dilations. After this coherent evolution, the environment is traced out and the branches are coherently recombined (control post-selection and trace out). Lastly, the effectively evolved system undergoes the second measurement. Time runs from left to right.}
\label{fig:supshared}
\end{figure*}

We now introduce the scheme of coherently controlling the choice of equivalent Stinespring dilations (see Fig.\ref{fig:supshared}). Here, the control qubit coherently selects between two unitarily equivalent dilations of the same noise channel. The two dilations are related solely by an environment-only unitary transformation:
\begin{equation} 
W_{D} = \ket{0}\!\bra{0}_C \otimes U_{0} + \ket{1}\!\bra{1}_C \otimes U_{1}, 
\label{eq:cunit}
\end{equation} 
with $U_{0} = U_{SE}$, $U_{1} = U'_{SE}$, where
\begin{equation} 
U'_{SE} = (\mathbb{I}_S \otimes V_E)\, U_{SE}, 
\end{equation} 
with $V_E$ being an arbitrary environmental unitary. Here, `D' denotes the case of coherent control of dilations.  Since the environment is initialized in a fixed state and $V_E$ acts only on the environment, both dilations induce identical action of the  channel on the system after tracing out the environment.

Note here that unlike the independent-environment construction discussed above, the two states of the control do not correspond to interaction with two different environmental registers. Instead, both the states control the action on the same system-environment Hilbert space $(S \otimes E)$, but implement two distinct yet unitarily equivalent Stinespring dilations of the same reduced noise channel.

\begin{equation}
\operatorname{Tr}_{E}\left[ U_{0} \left( \rho_S\otimes \rho_E \right) U_{0}^{\dagger} \right] 
= \operatorname{Tr}_{E} \left[ U_{1} \left( \rho_S\otimes \rho_E \right) U_{1}^{\dagger} \right] 
= \mathcal{E}(\rho_S), 
\label{eq:equivdila}
\end{equation}

\emph{Joint state and conditional system states:} Consider now the initial joint state of the control, system, and the environment to be
\begin{equation}
\rho^{D}_{\mathrm{tot}}(0) =
\rho_C(0)\otimes \rho_S(0) \otimes |0\rangle\!\langle0|_{E},
\end{equation}
with $\rho_C(0) = |q\rangle\!\langle q|{_C}$, where `q' can be either 0 or 1.\\

The controlled unitary on the combined system of the control qubit $C$, the system $S$ and the environment $E$ (see Eq. \ref{eq:cunit}) evolves the total state to
\begin{equation}
\rho^{D}_{\mathrm{tot}}(\theta) =
W_{D} \rho^{D}_{\mathrm{tot}}(0) W_{D}^\dagger,
\end{equation}
from which the reduced control-system state is obtained as
\begin{eqnarray}
\rho^{D}_{CS}
&=& \operatorname{Tr}_{E} \!\left[ \rho^{D}_{\mathrm{tot}} \right] \nonumber \\
&=& \ket{q}\bra{q}_{C} \otimes \mathcal{E}_{q}(\rho_S), 
\end{eqnarray}
where $\mathcal{E}_{q}(\rho_S)$ is the channel realized through interaction with the environment through the specific dilation selected by the value of `q'. Thus, when the control C is initialized in the state $\ket{q}$, the system S undergoes the reduced dynamics induced through the corresponding dilation $U_{q}$. Consequently, if C is prepared in a maximally incoherently state, i.e., $\rho_C(0)~=~ \frac{(\ket{0}\bra{0}_{C}~+~\ket{1}\bra{1}_{C})}{2}$, the system S evolves either through the specific unitary dilation $U_{0}$ or $U_{1}$ with probability $0.5$. 

Since both the dilations realize the same reduced channel, the system state simply reduces to $\mathcal{E}(\rho_S)$. As such, incoherent control of the two equivalent Stinespring dilations is operationally indistinguishable from the use of single deterministic noise channel.\\

\emph{Reduced channel and cross maps:}
Now, let us define the reduced channel on $S$
\begin{equation}
\mathcal{E}_{ii}(\rho_S) :=
\operatorname{Tr}_E\!\Big[U_{i} (\rho_S\otimes|0\rangle\!\langle 0|_{E})U_{i}^\dagger\Big],
\end{equation}
and the cross maps
\begin{equation}
\begin{aligned}
\Gamma_{01}(\rho_S):=&\operatorname{Tr}_E\!\Big[U_{0}(\rho_S\otimes|0\rangle\!\langle0|_{E})U^{\dagger}_{1}\Big],\\
\qquad
\Gamma_{10}(\rho_S):=&\operatorname{Tr}_E\!\Big[U_{1}(\rho_S\otimes|0\rangle\!\langle0|_{E})U_{0}^{\dagger}\Big].
\end{aligned}
\end{equation}
These cross maps are linear but are generally not CPTP.\\

Now, consider the input state of the control qubit to be, $\rho_C(0)=|\pm\rangle\langle\pm|_{C}$ with $|\pm\rangle_{C}=(|0\rangle_{C}\pm|1\rangle_{C})/\sqrt2$.
After applying $W_{D}$ and tracing out $E$, the joint state becomes

\begin{equation}
\begin{aligned}
\rho^{D}_{CS} =& \frac{1}{2} \Big(|0\rangle\!\langle0|_{C} \otimes \mathcal{E}_{00}(\rho_S) + |1\rangle\!\langle1|_{C} \otimes \mathcal{E}_{11}(\rho_S)\Big) \\
&\pm \frac12 \Big(|0\rangle\!\langle1|_C \otimes \Gamma_{01} (\rho_S) + |1\rangle\!\langle0|_C \otimes \Gamma_{10} (\rho_S) \Big),
\end{aligned}
\label{eq:rhoCS_dila}
\end{equation}

Equivalently, analogous to the independent environment case, we can define the \textit{effective} system–environment operators associated with the two control outcomes $\pm$ as,

\begin{equation}
A_{\pm}^{D} = \frac{1}{2}  \left( U_{0}\pm U_{1} \right).
\end{equation}
Notice here that the operators $U_0$, $U_1$ act on the joint Hilbert space $(S \otimes E)$ unlike the case discussed above.
The corresponding unnormalized \textit{conditional} channel can then be written as,
\begin{equation}
\label{eq:EDS}
\widetilde{\mathcal{E}}_{\pm}^{D}(\rho_S)
= \operatorname{Tr}_{E} \left[A_{\pm}^{D}
\left(\rho_S\otimes |0\rangle\langle 0|_{E} \right) A_{\pm}^{D\dagger} \right].
\end{equation}

Expanding Eq.~(\ref{eq:EDS}), we get,
\begin{align}
\widetilde{\mathcal{E}}_{\pm}^{D}(\rho_S)
&= \frac{1}{4} \left[ \mathcal{E}_{00}(\rho_S)+\mathcal{E}_{11}(\rho_S) \right] \nonumber\\
&\quad \pm \frac{1}{4} \left[ \Gamma_{01}(\rho_S) + \Gamma_{10}(\rho_S) \right].
\label{eq:dila-map}
\end{align}
Since $U_0$, $U_1$ are unitarily equivalent Stinespring dilation of the same noise channel, we have 
\begin{equation}
\mathcal{E}_{00}(\rho_S)= \mathcal{E}_{11}(\rho_S) = \mathcal{E}(\rho_S),
\end{equation}
implying that the deterministic (reduced) action is identical leading to,
\begin{equation}
\begin{aligned}
\rho^{D}_{CS}  =&
\frac{\mathbb{I}_C}{2}\otimes \mathcal{E}(\rho_S)\\
\pm&
\frac{1}{2}\Big(
|0\rangle\langle1|_C\otimes \Gamma_{01}(\rho_S)
+
|1\rangle\langle0|_C\otimes \Gamma_{10}(\rho_S)
\Big).
\end{aligned}
\end{equation}
Now, measuring $C$ in the Hadamard basis $\{|\pm\rangle_{C}\}$ basis with projectors $\Pi_{C,\pm}=|\pm\rangle\langle\pm|_{C}$
yields the unnormalized conditional system states,
\begin{equation}
\begin{aligned}
\tilde{\rho}^{D}_{S,\pm} =&
\operatorname{Tr}_C\!\big[(\Pi_{C,\pm}\otimes\mathbb{I}_S)\rho^{D}_{CS}\big]\\
=& \frac12\,\mathcal{E}(\rho_S) \pm
\frac{1}{4}\Big(\Gamma_{01}(\rho_S) + \Gamma_{10}(\rho_S)\Big).
\label{eq:conditional_dila}
\end{aligned}
\end{equation}
Equivalently, the unnormalized conditional channel reduces to the form, 
\begin{equation}
\widetilde{\mathcal{E}}_{\pm}^{D}(\rho_S)
= \frac{1}{2}\mathcal{E}(\rho_S) \pm \frac{1}{4}\Big(\Gamma_{01}(\rho_S) + \Gamma_{10}(\rho_S)\Big).
\label{eq:identical-dila}
\end{equation}
The corresponding post-selection probabilities are as follows:
\begin{equation}
P^{D}_\pm=\operatorname{Tr}\tilde{\rho}{^{D}}_{S,\pm},
\end{equation}
 and the normalized conditional states are given by
\begin{equation}
\rho{{^{D}}_{S,\pm}} = \frac{\tilde{\rho}{^{D}}_{S,\pm}}{P{^{D}}_\pm}.
\end{equation}

\begin{figure}[!htbp]
\includegraphics[width=0.5\textwidth]{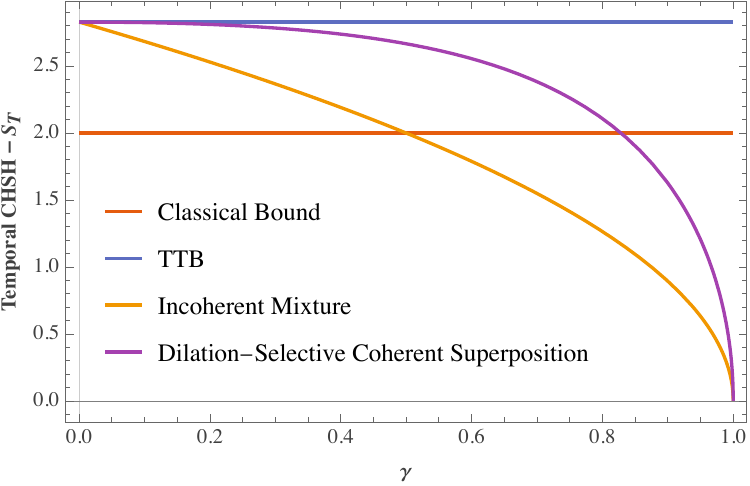}
\caption{A plot of the variation in the temporal CHSH inequality as a function of $\gamma$ is shown. The orange curve represents the case when Stinespring dilations are incoherently chosen. The purple curve is due to coherent selection of the unitarily equivalent dilations. Notice that the violation of the temporal CHSH inequality shows activation beyond the ADC threshold and up to $\gamma \approx 0.83$.}
\label{fig:dila-tchsh}
\end{figure} 

\emph{Specialization to ADC:}
For the ADC dilation with $\rho_E~=~|0\rangle\langle0|$, the dilations are $U_0=U_{SE}(\theta)$ and $U_1~=~U^{-1}_{SE}(\theta)~=~ U^\dagger_{SE}(\theta)~=~U_{SE}(-\theta)$. The set of Kraus operators corresponding to the unitaries $U_{SE}(\theta)$, $U^{-1}_{SE}(\theta)$ are $\{K_0(\theta), K_1(\theta)\}$ and $\{K_0(\theta), -K_1(\theta)\}$ respectively. Since the sign of a Kraus operator does not affect the reduced channel, both realize the same ADC.

To this end, recognizing that $\mathcal{E}_{00}(\rho_S)=\mathcal{E}_{11}(\rho_S) = \mathcal{E}_{\theta}(\rho_S)$ and $\Gamma_{01}(\rho_S)=\Gamma_{10}(\rho_S) = \Gamma_{\theta}(\rho_S)$, 
we can easily express,
\begin{equation}
\begin{aligned}
\mathcal{E}_{\theta}(\rho_S)=&K_0(\theta)\rho_S K_0^\dagger(\theta) + K_1(\theta)\rho_S K_1^\dagger(\theta),\\
\Gamma_\theta(\rho_S)=&K_0(\theta)\rho_S K_0^\dagger(\theta) - K_1(\theta)\rho_S K_1^\dagger(\theta).
\end{aligned}
\end{equation}

As such,
\begin{equation}
\begin{aligned}
\tilde{\rho}^{D}_{S,\pm}(\theta)
=& \frac12\,\mathcal{E}_\theta(\rho_S)\\
&\pm \frac{1}{2}\Big(K_0(\theta)\rho_S K_0^\dagger(\theta) - K_1(\theta)\rho_S K_1^\dagger(\theta)\Big).
\end{aligned}
\end{equation}
Hence, $\tilde{\rho}{^{D}}_{S,+} = K_0(\theta)\rho K_0^\dagger(\theta)$ and $\tilde{\rho}{^{D}}_{S,-} = K_1(\theta)\rho K_1^\dagger(\theta)$. This scenario corresponds to the effective evolution due to the coherent superposition of the joint unitary $U_0=U_{SE}(\theta)$ and its inverse $U_1 = U^{-1}_{SE}(\theta) = U^\dagger_{SE}(\theta)= U_{SE}(-\theta)$ generated by the reverse Hamiltonian `$-H$' \cite{CKM+25}. Notice that the measurement of the control (in the coherence preserving basis) in this scenario  coherently separates the action of $K_0$ and $K_1$, whereas the updated state in the selected $+$ branch in the independent environment implementation is a mixture of the actions of the Kraus operators $K_0$ and $K_1$. 

Here again, we now have two clear ways to check activation in dilation-level coherent control scenario:
\begin{enumerate}
    \item Prepare the control in a maximally incoherent mixed state, i.e., $\rho_C(0)= \frac{(\ket{0}\bra{0}_{C} + \ket{1}\bra{1}_{C})}{2}$. 
\end{enumerate}
    \noindent Analogous to the corresponding scenario in the case of coherent control of superposition arising due to independent environments, here too the control only implements a classical random selection between the two unitarily equivalent Stinespring dilations. Since both the equivalent dilations lead to the same reduced channel on the system, the updated state of the system is identical to that obtained from a deterministic use of the single noise channel  when $\rho_0 = \frac{\mathbb{I}}{2}$,  with $P(a,b|A_i,B_j) =\frac{1}{2} \mathrm{Tr} \left[ M_{b|B_j} \mathcal{E}(M_{a|A_i})\right]$, from which we can infer that the incoherent mixing of the equivalent dilations does not activate the violation of temporal CHSH inequality beyond the usual deterministic ADC threshold.
    
\begin{figure} [!htbp]
\centering
\includegraphics[width=0.45\textwidth]{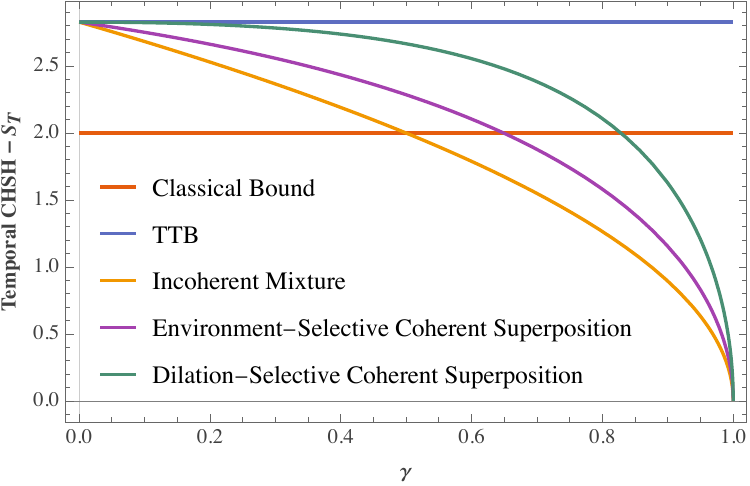}
\caption{The temporal CHSH inequality as a function of $\gamma$ under different considerations: (1) The orange curve denotes the case where the choice of the dilation is deterministic or incoherent (classical) mixture, (2) The purple is due to coherent control of the channels with independent environments and (3) The green corresponds to the coherent control of unitary equivalent dilations with the same environment.} 
\label{fig:hierarchy}
\end{figure}

\begin{enumerate}[resume]
    \item Prepare the control in the $\ket{+}_{C}$ state:
\end{enumerate}    
    \noindent Here too, after jointly evolving the system and the control, post-selecting the control before tracing it out, the two-time (unnormalized) probabilities can be calculated using the expression, 
    \begin{equation}
    \widetilde{P}(a,b|A_i,B_j) =\frac{1}{2} \mathrm{Tr} \left[ M_{b|B_j} \widetilde{\mathcal{E}}_{\pm}^{\mathrm{D}}(M_{a|A_i})\right].
    \label{eq:comprob}
    \end{equation} 
    The success probability of the post-selected control measurement is given by $P^{D}_\pm = \operatorname{Tr}\Big[\tilde{\rho}{^{D}}_{S,\pm}\Big]$ from which we can obtain the normalized two-time joint probability as $P(a,b|A_i,B_j) = \frac{\widetilde{P}(a,b|A_i,B_j)}{P^{D}_\pm}$. To this end, the interference terms in $\widetilde{\mathcal{E}}_{\pm}^{D}(M_{a|A_i})$ clearly aid in activation of the temporal CHSH inequality violation beyond the threshold obtained for incoherently controlled ADCs as well as the case of the coherent control implementation using independent environments (see Figs. \ref{fig:dila-tchsh}, \ref{fig:hierarchy} and  the Appendix \ref{app:cm} for the corresponding formulation in terms of the associated Choi state).

\section{Activation Thresholds}
\label{Sec:at}
We evaluate temporal CHSH inequality violations under both constructions using identical measurement settings (Observables in the $X-Y$ plane (see Eq. \eqref{eq:xymeas}).

The main findings are that there is clearly a distinct difference in the observation of the activation range pertaining to the violation of temporal CHSH inequality (see Fig.~\ref{fig:hierarchy}): In particular, 
\begin{itemize}
\item For coherent control of channels with independent environments, the expression for the temporal CHSH inequality becomes (see Appendix \ref{app:cm})
\begin{equation}
\mathcal{S}_{T}^\mathrm{ind} = \frac{8 \sqrt{2-2 \gamma }}{4 - \gamma},
\label{eq:tchshind}
\end{equation} 
and the success probability of post-selection corresponding to the $+$ control outcome is $P^\mathrm{ind}_+~=~\operatorname{Tr} \big[ \tilde{\rho}^\mathrm{ind}_{S,+} \big] = \frac{1}{2} + \frac{1}{4} (2 - \gamma)$. Consequently, the temporal CHSH inequality violation gets activated up to $\gamma_{c} \simeq 0.65$ ~(see Fig. \ref{fig:ind-tchsh}). 

\item Correspondingly, for the coherent control of unitarily-equivalent Stinespring dilations, the expression for the temporal CHSH inequality is (see Appendix \ref{app:cm}),
\begin{equation}
\mathcal{S}_{T}^{D} = \frac{4 \sqrt{2-2 \gamma }}{2 - \gamma},
\label{eq:tchshcom}
\end{equation}
and the success probability of post-selection corresponding to the $+$ control outcome is $P^\mathrm{D}_+ = \operatorname{Tr} \big[ \tilde{\rho}^\mathrm{D}_{S,+} \big] = 1 - \frac{\gamma}{2}$. To this end,
the violation can now persist up to $\gamma_c \simeq 0.83$ coinciding with the threshold reported  in \cite{KWS+21} using stochastic pre- and post- filtering operations~ (see Fig. \ref{fig:dila-tchsh}) for the maximum filter strength.
\end{itemize}

Thus, the coherent control at the dilation level significantly extends the activation region compared with the independent-environment implementation. The two thresholds are clearly separated  for the measurement settings considered here.

\section{Discussion}
\label{sec:dis}
In \cite{KWS+21}, it was shown that the degradation of temporal-CHSH inequality violation under amplitude damping can be revived using stochastic pre- and post- operations. Here we identify a physically distinct activation mechanism based on coherent control of equivalent system–environment Stinespring dilations. Although, for amplitude damping, the resulting temporal-CHSH behavior coincides with that obtained using the optimal filtering protocol, the underlying mechanisms are fundamentally different: in the present case, the revival arises from interference between alternative dilation-level realizations of the same channel, rather than from externally applied stochastic filters.

The enhanced activation observed in the equivalent dilation implementation can be traced to the fundamentally different role played by interference in the two coherently controlled constructions. In the independent-environment scenario, the two control branches couple the system to distinct environments. As a consequence, interference between the two channel implementation is limited by the presence of distinct environmental records and hence tracing out the environments is found to partially degrade the coherence between the controlled branches. The resulting post-selected dynamics can therefore enable the activation of the temporal CHSH inequality violation beyond the incoherent ADC threshold (which is identical to the simple application of the ADC), but only over a smaller range of the damping strengths.

In contrast, when the two interfering branches correspond to unitarily equivalent Stinespring dilations involving the same environment, the interference occurs directly at the level of the system--environment evolution before the environment is discarded. For the ADC dilations considered here, the resulting conditional evolution is therefore considerably richer and extends the regime in which temporal CHSH inequality violations can be observed. Importantly, this activation should not be interpreted as unconditional noise cancellation. The revival of the violation happens only from a conditioned ensemble corresponding to the particular quantum control outcome while the incoherently controlled, or the deterministic reduced dynamics remains noisy.  

An interesting consequence of our work contributes to the idea of employing temporal correlation experiments to witness analogous nonlocality-breaking property of noisy channels (see \cite{pal2015non, ku22, KWS+21} and Appendix \ref{app:cm}). Specifically, under initial setting-independent post-selection and employing the equatorial plane ($X-Y$ plane) measurements -- prescribed in  Eq.~\eqref{eq:xymeas} for the initial and later time measurements-- coherent control of equivalent Stinespring dilations activates the temporal CHSH inequality violation ($0.5 < \gamma \lesssim 0.83$). This matches the optimal hidden-nonmacrorealism activation threshold previously obtained using stochastic filtering in \cite{KWS+21}. Since the post-selection probability is independent of the initial measurement setting, the observed temporal-CHSH inequality violation certifies that the ADC is not a strongly nonlocality-breaking channel in this range. In contrast, the coherent control with independent-environments too activates the temporal-CHSH inequality violation and certifies the not strongly nonlocality-breaking property of ADC but only for a narrow  range ($0.5 < \gamma~ \lesssim 0.65$). 

In conclusion, our results demonstrate that coherent control can reveal operational differences between physical implementations of the same noisy channel. Although both constructions reduce to the same ADC when used deterministically, they give rise to distinct post-selected evolutions and consequently different temporal correlation structures. The observed activation of temporal nonclassicality therefore cannot be understood solely in terms of the underlying CPTP map. 

More broadly, these results align temporal nonclassicality tests with other scenarios in which coherent control makes different Stinespring realizations of the same quantum channel operationally distinguishable. In this setting, the choice of Stinespring dilation acquires direct operational significance: different dilations of the same channel, ordinarily regarded as equivalent from the perspective of reduced dynamics, can become distinguishable, when embedded within a coherently controlled protocol. This highlights a direct link between temporal quantum correlations, coherent control, and the physical realization of quantum processes.

\section{acknowledgements}
HSK, MP acknowledge support from NCN Poland, ChistEra-2023/05/Y/ST2/00005 under the project Modern Device Independent Cryptography (MoDIC). ASH acknowledges full support, while MP acknowledges that this work is partially carried out under IRA Programme, project no. FENG.02.01-IP.050006/23, financed by the FENG program 2021-2027, Priority FENG.02, Measure FENG.02.01., with the support of the FNP.

\bibliographystyle{apsrev4-2}
\bibliography{ref}

\appendix
\section{Choi Matrix and Non-locality breaking channels} 
\label{app:cm}
To understand the connection between temporal CHSH violations and
nonlocality-breaking property of the quantum channels, we consider it useful to employ the
Choi--Jamio{\l}kowski representation of quantum channels~\cite{Jamiolkowski1972,Choi1975,NielsenChuang}.

For a qubit channel
$\mathcal{E}$, the normalized Choi state is defined as
\begin{equation}
J_{\mathcal E} = (\mathbb I\otimes \mathcal E) \left(|\Phi^+\rangle\langle\Phi^+|\right),
\end{equation}
where $|\Phi^+\rangle = \frac{|00\rangle+|11\rangle}{\sqrt{2}}$ is a maximally entangled two-qubit state.  The inverse Choi--Jamio{\l}kowski relation is then given by  $\mathcal{E}(X) = 2\operatorname{Tr}_{A}  \left[ \left( X^{T}\otimes\mathbbm{1} \right) J_{\mathcal{E}} \right]$ from which one can reconstruct the channel using the elements of the Choi state. 

Temporal correlation function generated by a channel can then be expressed as expectation values evaluated on its Choi state. More specifically, using Alice's and Bob's observables, $A_i~=~\boldsymbol{\sigma}\cdot\boldsymbol{a}_i,$ $B_j~=~\boldsymbol{\sigma}\cdot\boldsymbol{b}_j$,
where $\|\boldsymbol{a}_i\| = \|\boldsymbol{b}_j\| = 1$, we have 
\begin{equation}
C_{ij} = \operatorname{Tr} \left[(A_i^T\otimes B_j)J_{\mathcal E}\right],
\label{eq:temp choi corr}
\end{equation}
Choosing a maximally mixed initial qubit state and projective measurements, the temporal correlations of the channel coincide mathematically with spatial correlations evaluated on the corresponding Choi state (with a transpose `$T$' appearing for Alice’s observable).
To this end, the temporal CHSH inequality associated with the channel
may now be expressed as
\begin{equation}
\mathcal{B}_T = A_0^T\otimes (B_0+B_1) +A_1^T\otimes (B_0-B_1)
\end{equation}
such that $\mathcal{S}_T = \operatorname{Tr}
\left[\mathcal{B}_T\, J_{\mathcal E}\right].$
This establishes a precise relation between temporal CHSH correlations and the spatial correlation obtained from the Choi state. Irrespective of the underlying physical realization, the mathematical form of these correlations is identical. As such, one can employ temporal quantum correlations for characterizing the nonlocality-breaking property of channels without invoking spatial quantum correlations \cite{ku22, KWS+21}.

Equivalently, thresholds for temporal CHSH violations can also be explored via the corresponding Choi states: The Choi matrix $J_{\mathcal E}$ for the ADC is given by 
\begin{equation}
J_{\mathcal E} =
\begin{pmatrix}
 \frac{1}{2} & 0 & 0 & \frac{\sqrt{1-\gamma}}{2} \\
 0 & 0 & 0 & 0 \\
 0 & 0 & \frac{\gamma}{2} & 0 \\
 \frac{\sqrt{1-\gamma}}{2} & 0 & 0 & \frac{1-\gamma}{2}
\end{pmatrix}
\label{ap:choi}
\end{equation}
and elements $t_{ij}$ of the two-qubit correlation matrix $T$ associated with the Choi matrix Eq.(\ref{ap:choi}) are evaluated using 
\begin{equation}
t_{ij} = {\rm Tr} \left[ J_{\mathcal E} (\sigma_i\otimes\sigma_j) \right],
\qquad i,j = \{1,2,3\}.
\end{equation}
Recall that a generic two-qubit state can be expressed as 
\begin{align*}
\rho_{AB} = \frac{1}{4} \bigg[\mathbb{I}_2\otimes \mathbb{I}_2
+ \sum_{i=1}^{3} a_i \sigma_i\otimes \mathbb{I}_2\\
+ \sum_{j=1}^{3} b_j \mathbb{I}_2\otimes \sigma_j
+& \sum_{i,j=1}^{3} t_{ij}\,\sigma_i\otimes\sigma_j \bigg],
\end{align*}
where $\textbf{a}, \textbf{b} \in \mathbb{R}^{3}$ are the local Bloch vectors and $\textbf{T} \in \mathbb{R}^{3\times3}$ is the correlation matrix constructed from the $t_{ij}, ~ i,j = 1,2,3~ \rm{and} ~(\sigma_1, \sigma_2, \sigma_3)$ are the Pauli matrices.
The maximal temporal CHSH value is then given by~\cite{horodecki1995violating}
\begin{equation}
\mathcal{S_T} = 2\sqrt{t^2_1+t^2_2}.
\end{equation}
where $t^2_1$ and $t^2_2$ correspond to two highest eigenvalues of the matrix $T^T\,T.$
This leads to the two-qubit correlation matrix associated with the ADC Choi matrix as,
\begin{equation}
\label{ap:tADC}
T_\mathrm{ADC}=\mathrm{diag}(\sqrt{1-\gamma},\, -\sqrt{1-\gamma},\, (1-\gamma)),
\end{equation}
 and therefore $t_1^2=t_2^2=1-\gamma,$
resulting in $S_T^{\max} = 2\sqrt{2(1-\gamma)}.$
In other words, the temporal CHSH inequality is violated whenever
\begin{equation}
2\sqrt{2(1-\gamma)}>2 \Rightarrow \gamma_c=0.5.
\end{equation}

In the spatial setting, a two-qubit state may fail to violate the CHSH inequality and yet reveal \emph{hidden} CHSH nonlocality after suitable \emph{local-filtering} operations. The Choi state of the amplitude-damping channel possesses \emph{hidden} CHSH nonlocality~\cite{verstraete2002hidden, pal2015non} for $\frac{1}{2} < \gamma < 2(\sqrt{2}-1)\approx 0.83$.  

In the temporal scenario studied here, coherent control together with post-selection generates effective conditional maps that are completely positive but generally not trace preserving. Their temporal-CHSH properties can be analyzed through the corresponding Choi operators. In particular, if the normalized Choi operator associated with a setting-independent post-selected implementation violates the CHSH inequality, then the corresponding channel cannot be strongly CHSH nonlocality-breaking. Hence, temporal CHSH activation provides a channel-based certification of hidden CHSH nonlocality without requiring a spatial correlation (Bell) experiment.

\paragraph{Independent-environments.}
For the independent-environment implementation, the normalized postselected Choi matrix is given by 
\begin{equation}
J^{\mathrm{ind}}_{+}
=
\begin{pmatrix}
 \frac{2}{4 - \gamma} & 0 & 0 & \frac{2 \sqrt{1-\gamma }}{4 - \gamma} \\
 0 & 0 & 0 & 0 \\
 0 & 0 & \frac{\gamma }{4-\gamma } & 0 \\
 \frac{2 \sqrt{1-\gamma }}{4 - \gamma} & 0 & 0 & \frac{2(1 - \gamma)}{4 - \gamma} \\
\end{pmatrix}
\label{eq:+choiind}
\end{equation}
and the corresponding two-qubit correlation matrix is evaluated as:
\begin{equation}
T_{\mathrm{ind}}
= \mathrm{diag} \!\left( \frac{\sqrt{1-\gamma}}{1-\gamma/4},
-\frac{\sqrt{1-\gamma}}{1-\gamma/4},
\frac{4 - 3\gamma}{4 - \gamma} \right).
\label{eq:+Tind}
\end{equation}

Two largest eigenvalues of $T_{\mathrm{ind}}^TT_{\mathrm{ind}}$ are found to be 
\begin{equation}
t_1^2=t_2^2 = \frac{1-\gamma}{(1-\gamma/4)^2} = 
\frac{16(1-\gamma)} {(4-\gamma)^2},
\end{equation}
from which we get, $\mathcal{S}_T^{\max, \mathrm{ind}}=2\sqrt{t_1^2+t_2^2} = \frac{8 \sqrt{2-2 \gamma }}{4 - \gamma}$. Notice that this matches with the Eq.~\eqref{eq:tchshind} of the main text.
From this, the threshold value of damping strength is found to be 
\begin{equation}
\gamma_c \approx 0.65.
\end{equation}
below which the (spatial) CHSH inequality is not violated. Equivalently, the conditional channel associated with the post-selected coherent control of independent-environment implementation exhibits temporal CHSH inequality violation beyond $\gamma = 0.5$ all the way up to $\gamma \approx 0.65.$ 

\paragraph{Unitarily equivalent Stinespring dilations.}
In the case of  coherent control of unitarily equivalent Stinespring dilations, conditioning results in the normalized Choi matrix: 
\begin{equation}
J^{D}_{+}=
\begin{pmatrix}
 \frac{1}{{2-\gamma}} & 0 & 0 & \frac{\sqrt{1-\gamma}}{{2-\gamma}} \\
 0 & 0 & 0 & 0 \\
 0 & 0 & 0 & 0 \\
 \frac{\sqrt{1-\gamma}}{{2-\gamma}} & 0 & 0 & \frac{1-\gamma}{{2-\gamma}}
\end{pmatrix}.
\label{eq:+choicom}
\end{equation}
The associated two-qubit correlation matrix is given by,
\begin{equation}
T_D = \mathrm{diag} \!\left(\frac{2\sqrt{1-\gamma}}{2-\gamma},
-\frac{2\sqrt{1-\gamma}}{2-\gamma}, 1 \right).
\label{eq:+Tcom}
\end{equation}
Substituting the highest eigenvalues $t_1^2=1,$ $t_2^2=\frac{4(1-\gamma)}{(2-\gamma)^2}$ of the matrix $T_D^T\,T_D$ in 
$\mathcal{S}_T^{\max}=2\sqrt{t_1^2+t_2^2}$, we find that violation of the temporal CHSH inequality is observed whenever, 
\begin{equation}
1 + \frac{4(1-\gamma)}{(2-\gamma)^2} > 1.
\label{eq:full}
\end{equation}
This implies that the CHSH inequality violation extends from $\gamma =0.5$ (incoherent mixture scenario which is identically equal to simply the deterministic use of ADC) to the entire range of $\gamma$.
To this end, a clean nonlocality-breaking channel question  as far as the temporal correlation experiment is concerned is the following: \emph{Can a temporal correlation test identify nonlocality-breaking property of a quantum channel without relying on initial measurement setting?} To answer this, we employ equatorial plane ($X-Y$ plane) measurements (see Eq.~\eqref{eq:xymeas}) for both the initial and later time measurements. With this restriction on the measurement settings, we find that coherent control of equivalent Stinespring dilations matches the optimal hidden-nonmacrorealism activation previously obtained using stochastic filtering \cite{KWS+21}, i.e., $0.5 < \gamma < 0.83$, while the coherent control with independent environments yields a strictly smaller activation region ($0.5 < \gamma < 0.65$). 

Even though the coherent control of equivalent dilations results in a better activation range, we observe that, unlike the corresponding normalized conditional Choi matrix  (Eq.~\ref{eq:+choicom}), the temporal CHSH inequality violation is activated only up to  $\gamma_c \approx 0.83$ and not for the entire range of $\gamma$. In order to match the CHSH expression (Eq.~\ref{eq:full}) corresponding to the conditional Choi matrix (see Eq.~\ref{eq:+choicom}), one has to employ measurements in the $X-Z$ plane. However, doing so would lead to non-uniform (setting-dependent) post-selection of the controlled implementation of the system-environment unitary interactions. 
 
\end{document}